\documentclass[review,12pt]{elsarticle}
\usepackage{amssymb}
\usepackage{amsthm}
\usepackage{framed} % To put a box to the nomenclature
\usepackage{amsmath,color}
\usepackage{mathrsfs}
\usepackage{graphicx}
\usepackage{epstopdf}
\usepackage{float}
\usepackage{caption}
\usepackage{subcaption}
\usepackage{bm}
\usepackage{bbm}
\usepackage{upgreek} %For non-italic greek letters
\usepackage{mathrsfs}
\usepackage{cleveref}
\usepackage{siunitx}
\usepackage{xcolor}
\usepackage[margin=2cm]{geometry}% TO PLAY WITH THE MARGINS
\usepackage{soul}
\usepackage{accents}

\usepackage{tikz}
\usetikzlibrary{patterns}
\usetikzlibrary{arrows.meta}

\usepackage{color,soul} %to highlight text
\usepackage{color} %To highlight references in the text
\usepackage{bm}%To apply bulky font to greek letters
\biboptions{sort&compress}
\soulregister\citep7 % To allow citep to be inside of highlighted text
\soulregister\citet7 % Idem with citet
\soulregister\citealp7 % Idem with citealp
\newsavebox{\measurebox} %To create a 1 column figure with a 2 column figure on its side
\usepackage{titlesec} %To use paragraph as subsubsubsection
\usepackage[T1]{fontenc} % This is to convert the paper easily to Word
\usepackage{lmodern} % This is to convert the paper easily to Word
\input{glyphtounicode}  % This is to convert the paper easily to Word

\def\onedot{$\mathsurround0pt\ldotp$}
\def\cdddot#1{% three dots 
  \mathbin{\vcenter{\baselineskip.67ex
    \hbox{\onedot}\hbox{\onedot}\hbox{\onedot}%
  }}%
}

\journal{Extreme Mechanics Letters}

\makeatletter
\def\@author#1{\g@addto@macro\elsauthors{\normalsize%
    \def\baselinestretch{1}%
    \upshape\authorsep#1\unskip\textsuperscript{%
      \ifx\@fnmark\@empty\else\unskip\sep\@fnmark\let\sep=,\fi
      \ifx\@corref\@empty\else\unskip\sep\@corref\let\sep=,\fi
      }%
    \def\authorsep{\unskip,\space}%
    \global\let\@fnmark\@empty
    \global\let\@corref\@empty  %% Added
    \global\let\sep\@empty}%
    \@eadauthor={#1}
}
\makeatother

\titleformat{\paragraph}
{\normalfont\normalsize\itshape}{\theparagraph}{1em}{}
\titlespacing*{\paragraph}
{0pt}{3.25ex plus 1ex minus .2ex}{1.5ex plus .2ex}

\begin{document}

\begin{frontmatter}



\title{Fracture of bio-cemented sands}


\author{Charalampos Konstantinou\fnref{UCY}}
%\ead{ck494@cam.ac.uk}

\author{Emilio Mart\'{\i}nez-Pa\~neda\corref{cor1}\fnref{IC,Oxf}}
\ead{e.martinez-paneda@imperial.ac.uk}

\author{Giovanna Biscontin\fnref{Cam}}

\author{Norman A. Fleck\fnref{Cam}}

\address[UCY]{Department of Civil and Environmental Engineering, University of Cyprus, 1678 Nicosia, Cyprus}

\address[IC]{Department of Civil and Environmental Engineering, Imperial College London, London SW7 2AZ, UK}

\address[Cam]{Department of Engineering, Cambridge University, CB2 1PZ Cambridge, UK}

\address[Oxf]{Department of Engineering Science, University of Oxford, Oxford OX1 3PJ, UK}

\cortext[cor1]{Corresponding author.}

\begin{abstract}
Bio-chemical reactions enable the production of biomimetic materials such as sandstones. In the present study, microbiologically-induced calcium carbonate precipitation (MICP) is used to manufacture laboratory-scale specimens for fracture toughness measurement. The mode I and mixed-mode fracture toughnesses are measured as a function of cementation, and are correlated with strength, permeability and porosity. A micromechanical model is developed to predict the dependence of mode I fracture toughness upon the degree of cementation. In addition, the role of the crack tip $T$-stress in dictating kink angle and toughness is determined for mixed mode loading. At a sufficiently low degree of cementation, the zone of microcracking in the vicinity of the crack tip is sufficiently large for a crack tip $K$-field to cease to exist and for crack kinking theory to not apply. The interplay between cementation and fracture properties of sedimentary rocks is explained; this understanding  underpins a wide range of rock fracture phenomena including hydraulic fracture.\\
\end{abstract}

\begin{keyword}

Bio-cementation \sep Fracture \sep Cemented sands \sep bio-mediated materials \sep microbiologically-induced calcium carbonate precipitation


%EML is a letter-sized journal. We expect your submission to be no more than 4,000 words and up to 5 figures if it is a regular journal paper. EML also publishes review articles and no word limit is applied to reviews. (NOTE: MOST PAPERS HAVE MORE THAN 5 FIGs)

\end{keyword}

\end{frontmatter}



\section{Introduction}
\label{Sec:Introduction}

There is a growing interest in using our understanding of the natural environment to inspire the invention of new materials and thereby expand material property space \cite{Coyle2018,Liu2020a}. This includes the development of \emph{bio-mimetic} materials \cite{Bauer2015,Da2020}, and the use of naturally occurring bio-chemical processes to create \emph{bio-mediated} materials with unique properties \cite{Deuerling2018}. Bio-cementation techniques have emerged as an environmentally-friendly approach to increase the strength of soils \cite{Chu2012}, heal cracks in concrete \cite{Lors2017}, and manufacture low-carbon building materials \cite{Iqbal2021}. Among existing bio-cementation techniques, microbiologically induced calcium carbonate precipitation (MICP) has attracted particular attention for its ability to control the degree of cementation in the production of bio-cemented sands that mimic natural sandstones \cite{Konstantinou2021}. \\

\begin{figure}[H]
\begin{center}
\includegraphics[width=1\textwidth]{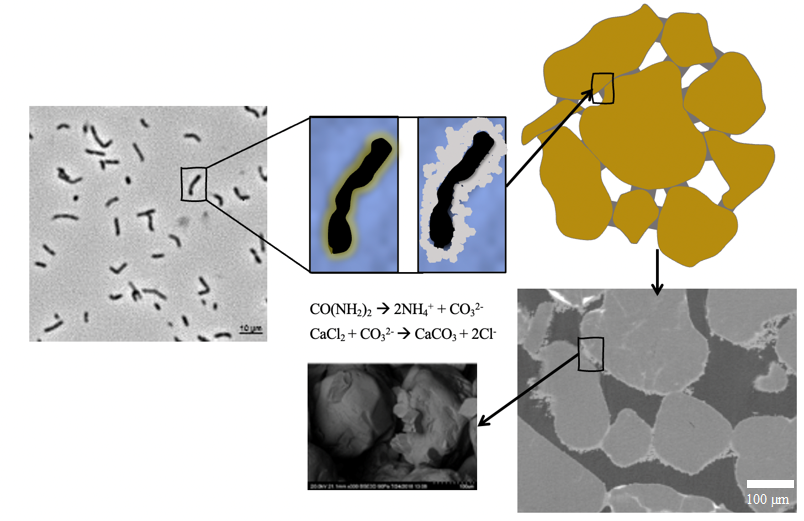}
\caption{The use of bacteria to fabricate cemented sands. A sketch of  microbially-induced calcium carbonate precipitation (MICP).}
\label{fig:Sketchmicpprocess}
\end{center}
\end{figure}

MICP involves the distribution and settlement of urease-producing bacteria within a granular matrix, see Fig. \ref{fig:Sketchmicpprocess}. First, a suspension of bacteria are introduced into the medium. The bacteria settle near the contact points between sand particles, and the cementation liquid is introduced repeatedly to generate calcium carbonate that binds the particles together. The cementation liquid consists of urea and a calcium source, usually calcium chloride. Calcium cations are attracted to the negatively charged cell walls of the bacteria and the microbes hydrolise urea to produce carbonate anions. Calcium carbonate precipitates on the bacterial wall, and consequently the position of the calcium carbonate crystals within the granular medium coincides with the location of the bacterial cells at contact points between the particles of a fine-grained sand \cite{Konstantinou2021a}. The mass of cement between particles correlates with the concentration of cementation solution and this provides a means of generating specimens with selected levels of cementation. The above strategy has recently been exploited in studies on the tensile and compressive strength of artificial bio-mediated sandstones \cite{Feng2016,Cui2017,Nafisi2019,Konstantinou2021b}, and was linked to the microscale response either via SEM or via acoustic emissions \cite{wang2021crackling}, providing suitable grounds to develop simulation-based approaches based on discrete and lattice element methods \cite{gago2022numerical,gong2023discrete,shen2023investigation,rizvi2019lattice,rizvi2019lattice1}. However, the fracture behaviour of bio-cemented sands remains unexplored despite its importance to phenomena such as hydraulic fracture, groundwater decontamination and grouting. \\

In the current study the fracture of artificially bio-cemented sandstones is investigated by combining experiment and modelling. The mode I and mixed-mode fracture responses are measured and are correlated with permeability, porosity, and tensile and compressive strengths of samples that have been manufactured using the same MICP recipe. The crack path under mixed-mode macroscopic loading is quantified, including the sensitivity of kink angle to the $T$-stress of the crack tip stress field. Test methods are detailed in Section \ref{Sec:Experiments}, and the experimental and modelling results are given in Section \ref{Sec:Results}.  A micromechanical model is developed to predict the dependence of toughness upon the degree of cementation (Section \ref{Sec:Model}). Concluding remarks are reported in Section \ref{Sec:Conclusions}.

%Since the mechanical behaviour of granular materials is primarily controlled by the degree of cementation \citet{Collins2009,Sitar1980,Saidi2005}, artificially cemented specimens of controlled properties can be generated with properties closely resembling those of the natural material.

\section{Experiments}
\label{Sec:Experiments}

The experimental protocol is now outlined, including the MICP parameters, sample preparation, and tests to measure mode I and mixed-mode fracture, permeability, porosity and bulk material strength (tensile and compressive). Silica sand was subjected to the MICP treatment; it comprised rounded to sub-rounded particles (medium sphericity) and a representative diameter ($D_{90}$) of 0.3 mm. The uniformity coefficient $D_{60}/D_{10}$ was 1.38, thereby classifying the granular medium as poorly graded (uniform distribution of particles). Cylindrical samples of diameter 70 mm and height 150 mm were first prepared, and then cut into smaller test coupons. Previous compressive strength measurements have shown good reproducibility but it should be noted that the scatter of the data increases with decreasing cementation level \cite{Konstantinou2021}.

\subsection{MICP parameters}

MICP allows for control of the degree of cementation and thereby the macroscopic mechanical properties. The treatment involved the sequential injection of the bacterial solution (BS) and the cementation solution (CS), as explained in the Introduction.  \\

\noindent (i) The bacterium \textit{Sporosarcina pasteurii} was used as its urease-synthesis behaviour can be controlled in a relatively straightforward manner. The growth medium comprised \SI{20}{g/L} yeast extract, \SI{10}{g/L} ammonium sulphate, \SI{20}{g/L} agar, and 0.13 M tris buffer. After 24 h of incubation at \SI{30}{\degreeCelsius}, the culture was harvested and stored at \SI{4}{\degreeCelsius}. The culture was introduced into liquid nutrient broth without agar, and placed in a shaking incubator for an additional 24 h to form the bacterial solution (BS); the BS was then introduced into sand columns. The optical density (OD), measured at a wavelength of 600~nm, OD$_{600}$, was between 1.5 and 2.0; the average urease activity was 0.8 (mM urea/h)/OD,  as measured by a conductivity assay on a BS diluted to an OD of 1.0 \citep{Whiffin2004}. \\

\noindent (ii) The cementation solution (CS) consisted of 0.375 M urea, 0.25 M calcium chloride ($\mathrm{CaCl}_2$), and 3 g/L nutrient broth. Previous studies have shown that this recipe gives an effective MICP treatment \cite{DeJong2006}. The combination of MICP parameters resulted in a uniform distribution of bacteria and chemicals without clogging. The rate of flow by gravity injection is balanced by the rate of MICP reaction, due to the chosen urease activity and concentration of the cementation solution. The long retention times promoted the efficient formation of calcium carbonate.\\

\subsection{Sample preparation and calcium carbonate content}

The injection direction in the cylindrical specimens was from top to bottom to give gravity-assisted permeation, thereby achieving uniform samples relative to other methods such as pumping. Before each injection, the excess solution was allowed to drain while the sample remained saturated. The time between two subsequent injections, the so-called retention time, was 24 hours. The degree of cementation was controlled by the number of CS injections, as the amount of calcium carbonate precipitation increases with the concentration of cementation solution, see Ref. \cite{Konstantinou2021} for additional details.\\

The calcium carbonate ($\mathrm{CaCO}_{3}$) content was measured according to the standard ASTM D4373-21, as follows. \SI{30}{mL} of hydrochloric acid (HCl) 2.5M was introduced into \SI{30}{g} of a dried and ground sample in order to generate carbon dioxide gas. Specifically, the $\mathrm{CaCO}_{3}$ reacts with HCl as follows:
\begin{equation}
    \mathrm{CaCO}_{3(s)} + \mathrm{2HCl}_{(aq)} \to \mathrm{CaCl}_{2(aq)}+ \mathrm{CO}_{2(g)} + \mathrm{H}_2 \mathrm{O}_{(l)} \label{react:feoh_feoh2}
\end{equation}
The quantity of calcium carbonate was calculated by measuring the mass of carbon dioxide released. To do so, the carbon dioxide gas was released into a calcimeter chamber and the gas pressure was measured using a pressure gauge. The gas pressure is related to the released amount of $\mathrm{CO}_{2(g)}$ and, from the stoichiomentry of the above reaction, the amount of calcium carbonate was calculated. The degree of cementation is reported  as the mass fraction of $\mathrm{CaCO}_{3}$. \\

Once the bio-cementation process had been completed, specimens of 70 mm diameter and 150 height were extracted from the molds. They were oven-dried at a temperature of 105$^{\circ}$C to remove all excessive soluble salts. The specimens were then cut into disks and a notch was introduced using a small knife.

\subsection{Fracture toughness tests}

\begin{figure}[H]
\begin{center}
\includegraphics[width=0.6\textwidth]{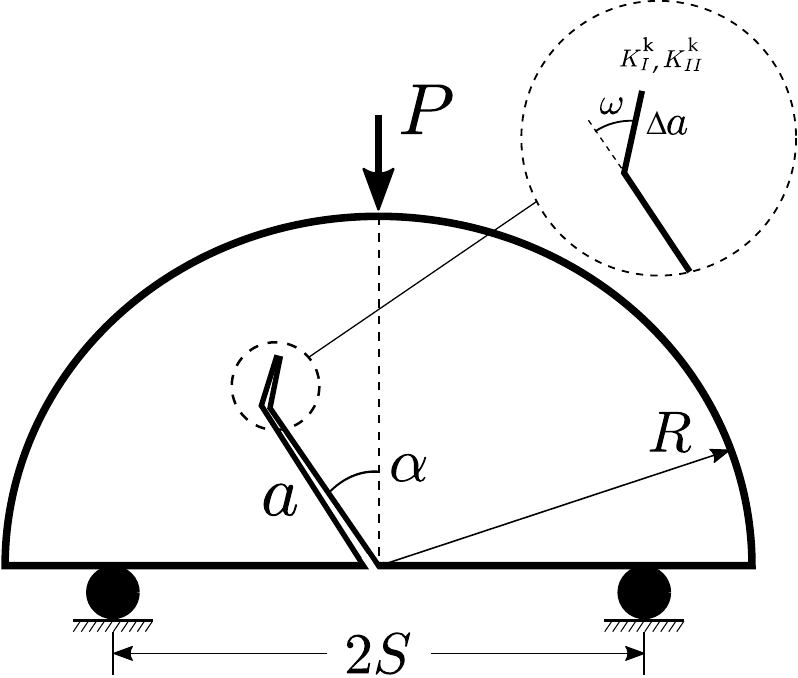}
\caption{The configuration of the mixed-mode fracture experiments, including the kink crack extension $\Delta a$ and kink angle $\omega$.}
\label{fig:FractureExpt}
\end{center}
\end{figure}

Three-point bend tests on semi-circular disks of radius $R=70$ mm were conducted to determine the mixed mode I and II fracture toughness, see Fig. \ref{fig:FractureExpt}. The test procedure followed the ISRM standard for fracture toughness measurement \cite{Kuruppu2014}. The samples were loaded quasi-statically under displacement-control at a rate of 0.05 mm/min. An initial notch of normalised length $a/R = 0.4$ was introduced, either parallel or inclined at an angle $\alpha$ with respect to the loading direction to give Mode I and Mixed-mode loading, respectively. The notch inclination angle $\alpha$ was of magnitude 0$^\circ$, 15$^\circ$, 30$^\circ$, 45$^\circ$, and 60$^\circ$ in order to span a wide range of mixed-mode behaviours. The span length is $S/R=1.6$. The mode I fracture toughness for the choice $\alpha=0$ is estimated from the peak load $P_{max}$ as
\begin{equation}
K_{Ic}=Y \frac{P_{max} \sqrt{\pi a}}{2Rt}
\end{equation}

\noindent where $t$ is the sample thickness and the calibration factor $Y$ is given by
\begin{equation}
Y=-1.297+9.516\frac{S}{2R} -\left(0.47+16.457\frac{S}{2R} \right) \frac{a}{R} + \left(1.071+34.401\frac{S}{2R}\right)\left( \frac{a}{R} \right)^2
\end{equation}

For mixed-mode loading, the mode I and mode II stress intensity factors $K_I$ and $K_{II}$, respectively, are related to the applied load $P$ by
\begin{equation}\label{eq:SIFs}
K_I = \frac{P \sqrt{\pi a}}{2Rt} Y_I \left( \alpha, \frac{a}{R}, \frac{S}{R} \right)  \,\,\,\,\,\,\,\,\,\,\,\,\,\, \text{and} \,\,\,\,\,\,\,\,\,\,\,\,\,\, K_{II} = \frac{P \sqrt{\pi a}}{2Rt} Y_{II} \left( \alpha, \frac{a}{R}, \frac{S}{R} \right)
\end{equation}

\noindent where the geometry factors $Y_I$ and $Y_{II}$ have already been obtained as a function of $\alpha$, $a/R$, and $S/R$ using finite element analysis \cite{Ayatollahi2007}. Similarly, the elastic $T$-stress \cite{Betegon1991} is given by
\begin{equation}
T = \frac{P}{2Rt} Y_T \left( \alpha, \frac{a}{R}, \frac{S}{R} \right) 
\end{equation}

\noindent where the calibration function $Y_T$ has alrady been reported in Ref. \cite{Ayatollahi2007}. 

\subsection{Relevant material properties}

Fracture toughness measurements were complemented by experiments aimed at correlating the degree of cementation of bio-cemented sands with relevant material properties: unconfined compressive strength (UCS), tensile strength, porosity, and permeability \cite{Konstantinou2021,Konstantinou2021a,Konstantinou2021b}. All tests were conducted on samples prepared using the same MICP protocol and over the same range of cementation ($\mathrm{CaCO}_{3}$ mass fraction) as used for the fracture toughness measurements.\\

Uniaxial compression tests were conducted on cylindrical samples by compressing them in the axial direction, in accordance with the standard ASTM D7012-14. The samples had a diameter of 70 mm and height of 150 mm, and the applied loading rate was 1.14 mm/min. The tensile strength was measured by means of the so-called Brazilian or indirect splitting test \cite{Navidtehrani2022a}, using disk samples of diameter 70 mm and thickness 30 mm, in accordance with the standard ASTM D2936. The relationship between porosity $p$ and $\mathrm{CaCO}_{3}$ mass fraction was obtained from the bulk dry density $\rho$ by using the usual relation:
\begin{equation}
p=1-\frac{\rho}{\rho_s}
\end{equation}

\noindent where $\rho_s$ is the weighted average density of silica and $\mathrm{CaCO}_{3}$  \citep{Konstantinou2021}. The permeability was measured by falling head tests, using the injection molds as rigid wall permeameters after all excess soluble salts had been removed by flushing with de-ionised water.

\section{Results}
\label{Sec:Results}

\subsection{Mode I fracture}
\label{Sec:ModeIresults}

\begin{figure}[H]
  \makebox[\textwidth][c]{\includegraphics[width=1.1\textwidth]{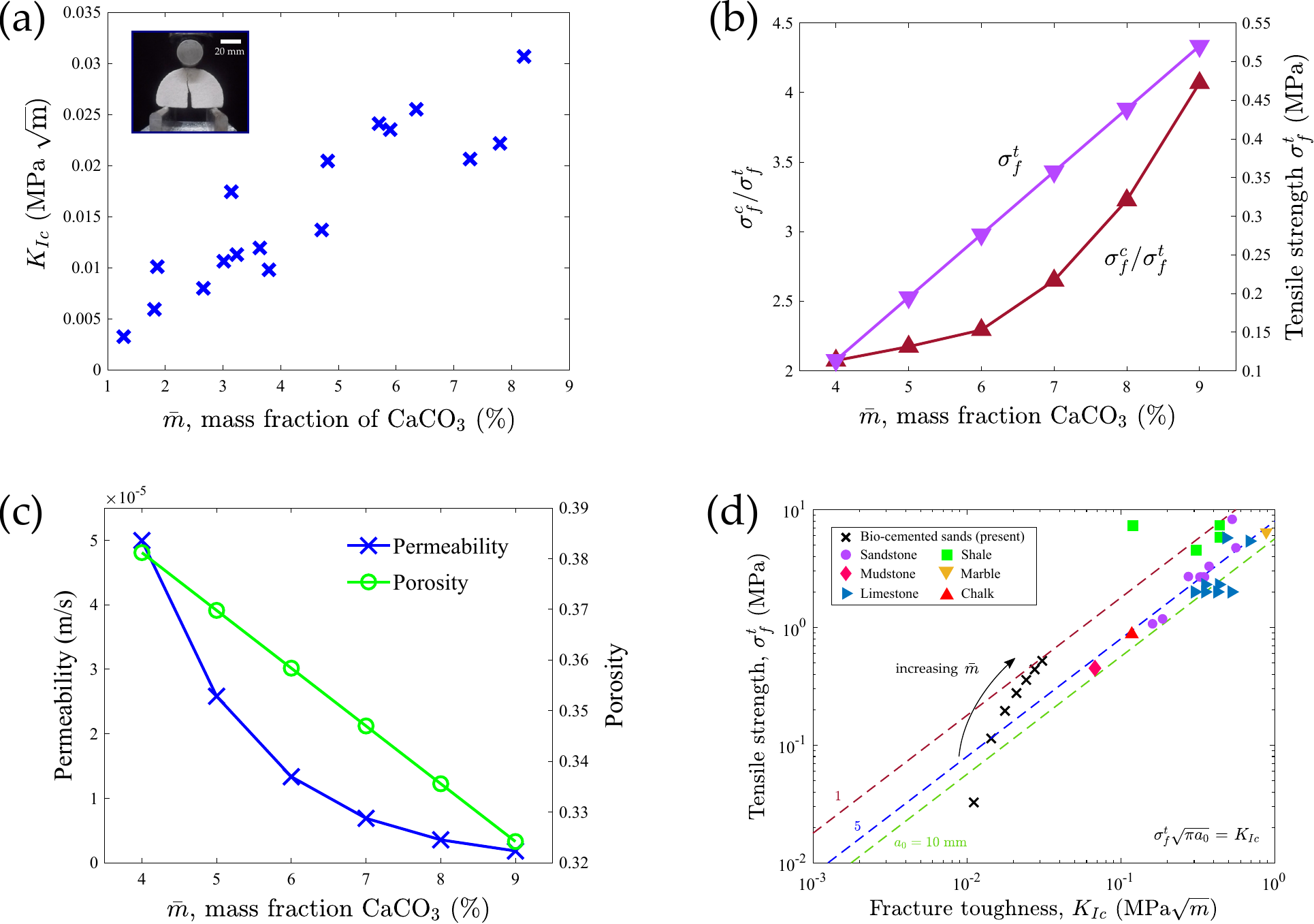}}%
  \caption{Experimental characterisation of the mechanical behaviour of bio-cemented sands: (a) Mode I fracture toughness $K_{Ic}$, (b) tensile strength $\sigma_f^t$ and compressive strength $\sigma_f^c$ as a function of cementation, (c) permeability and porosity as a function of cementation, and (d) correlation of tensile strength $\sigma_f^t$ to fracture toughness for a range of cemented rocks.}
  \label{fig:ExptModeI}
\end{figure}

The  dependence of mode I fracture toughness $K_{Ic}$ upon mass fraction $\bar{m}$ of $\mathrm{CaCO}_{3}$ is given in  Fig.\ref{fig:ExptModeI}a, while the tensile strength $\sigma_f^t$ and compressive strength $\sigma_f^c$ are plotted as a function of mass fraction $\bar{m}$  in  Fig. \ref{fig:ExptModeI}b. Note that both $K_{Ic}$ and $\sigma_f^t$ increase almost linearly with $\bar{m}$ whereas $\sigma_f^c$ scales with $\bar{m}$ in a non-linear fashion, consistent with the literature \cite{Terzis2018}. As the degree of cementation increases, such that $\bar{m}$ increases, the porosity and permeability both decrease, see  Fig. \ref{fig:ExptModeI}c. This behaviour is consistent with the notion that cementation fills pore space and leads to a drop in permeability and to an increase in both macroscopic strength and fracture toughness.  The range of porosities measured (0.32 to 0.37) is close to that exhibited by weak natural sandstones. A similar toughness-porosity relationship has been reported for sintered steels \cite{Fleck1981,Fleck1981b} due to the feature-in-common that discrete necks exist between particles. \\

It is expected that internal flaws exist in cemented rocks of characteristic length $2a_0$ (such as the diameter of a penny-shaped crack, or the length of a through-crack) which much exceeds the particle diameter $D$. Consequently, the tensile strength $\sigma_f^t$ is given by the criterion $K_I=\sigma_f^t \sqrt{\pi a_0}=K_{Ic}$ and the scatter in flaw size results in a scatter of tensile strength, as noted previously \cite{Konstantinou2021b}. It is instructive to plot the measured tensile strength versus fracture toughness for the bio-cemented sand of the present study in Fig. \ref{fig:ExptModeI}d, and to include published data \cite{GRANTA2021} for a range of cemented rocks. In addition, we make use of the relation between tensile fracture strength $\sigma_f^t$ and fracture toughness $K_{Ic}$, as given by $\sigma_f^t \sqrt{\pi a_0} = K_{Ic}$, to determine the critical defect size $a_0=(1/\pi)(K_{Ic}/ \sigma_f^t)^2$ that best fits the data. Taken together, the results show that bio-cemented sands have low values of both strength and fracture toughness relative to that of other rock-like materials. For the highest levels of cementation considered, bio-cemented sands can attain a tensile strength comparable to that of mudstone, but with a lower fracture toughness. The critical defect size $a_0$ that provides the best fit to the bio-cemented sands data decreases with increasing $\bar{m}$: low cemented bio-sands have critical defects as large as 5-10 mm, but as the degree of cementation increases $a_0$ is on the order of 1 mm. This is a smaller critical defect size than that of most rock-like materials and comparable to that of shale \cite{Kramarov2020}. We note in passing that the defect size $a_0$ in the cemented bio-sands and in naturally occurring cemented rocks \cite{GRANTA2021} is typically two orders of magnitude larger than the particle size $D$. Recall that the ASTM Standard E1820\footnote{Standard Test Method for Measurement of Fracture Toughness} demands that the crack length $a=28$ mm in the test program must exceed $2.5 (K_{Ic}/\sigma_f^t)^2=2.5 \pi a_0$ in order for the test to give a valid measure of fracture toughness. Consequently, fracture toughness measurements based on a crack length of $a=28$ mm are valid only for $a_0$ less than 3.5 mm. Thus, the $K_{Ic}$ tests are valid for $\bar{m} \geq 5\%$ by making use of Figs. \ref{fig:ExptModeI}a, \ref{fig:ExptModeI}b and \ref{fig:ExptModeI}d. This raises the immediate question of whether the bio-cemented sand behaves in the manner of an elastic-brittle solid when subjected to mixed mode loading.\\  

We further note that the critical defect size $a_0$ can be interpreted as the transition flaw size between strength-control and toughness-control. When a rock contains cracks much larger than $a_0$ its tensile strength is governed by its fracture toughness. In contrast, when small cracks are introduced (shorter than $a_0$), they have no effect upon the tensile strength of the rock as the tensile strength is set by the largest intrinsic flaw size of length $a_0$ \cite{gao2003materials1}.

\subsection{Mixed-mode fracture}

As shown in Fig. \ref{fig:FractureExpt}, the mixed mode experiments are conducted such that the initial crack is inclined at an angle $\alpha$ relative to the direction of applied load. Samples are produced for a wide range of $\alpha$ values and $\mathrm{CaCO}_{3}$ mass fractions. The direction of crack growth is characterised by the so-called kink angle $\omega$, see the inset of Fig. \ref{fig:FractureExpt}. The stress intensity factors that characterise the stress state at the tip of the kink crack ($K_I^k, K_{II}^k$) is related to the stress state at the tip of the original parent crack, as characterised by the stress intensity factors $K_I$ and $K_{II}$ \cite{Becker2001}. The $T$-stress associated with the parent crack additionally contributes to the stress intensity at the tip of a kink crack of finite length $\Delta a$. Following He and Hutchinson \cite{He1989} and He et al. \cite{He1991}, the stress intensity factors at the tip of the kink crack are given by
\begin{equation}\label{eq:K_Ik}
    K_I^k = A_I (\omega) K_I + B_I (\omega) K_{II} + C_I (\omega) T \sqrt{\Delta a}
\end{equation}
\begin{equation}\label{eq:K_IIk}
    K_{II}^k = A_{II} (\omega) K_I + B_{II} (\omega) K_{II} + C_{II} (\omega) T \sqrt{\Delta a}
\end{equation}

\noindent where the coefficients $A_i$, $B_i$ and $C_i$ depend upon the kink angle $\omega$ and are provided in Refs. \cite{He1989,He1991}.\\

\begin{figure}[H]
  \makebox[\textwidth][c]{\includegraphics[width=1.1\textwidth]{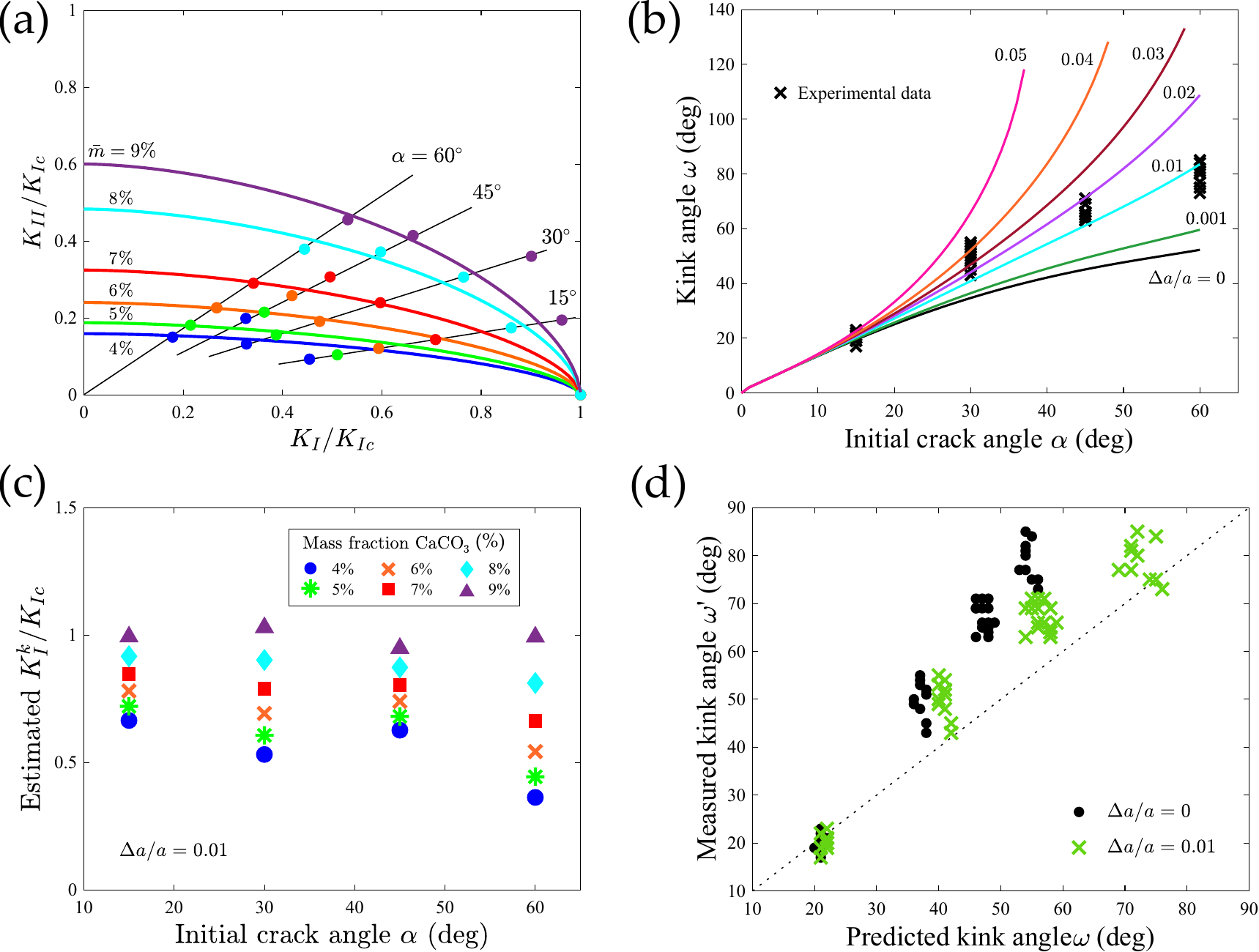}}%
  \caption{Mixed-mode fracture experiments: (a) critical stress intensity factors measured as a function of cementation ($\bar{m}$) and mode mixity (initial crack angle); (b) measured and predicted kink angles $\omega$ for the $K_{II}^k=0$ criterion upon considering different values of the $T$-stress coefficient, as characterised by $\Delta a /a=0.01$; (c) estimated values of $K_{I}^k$, normalised by $K_{Ic}$, for the choice of $\Delta a /a$ that provides the best fit to the measured kink angles; and (d) comparison between the measured and deduced kink angles for the choices $\Delta a = 0.01$ and $\Delta a = 0$, taking into account the role of the $T$-stress.}
  \label{fig:MixedModeExpt}
\end{figure}

The stress intensity factors ($K_I, K_{II}$) at the point of failure are determined from the critical applied load $P$, sample dimensions ($a/R, S/R$) and initial crack angle $\alpha$ via Eq. (\ref{eq:SIFs}). The locus of measured ($K_I, K_{II}$) at the initiation of crack growth  is generated for selected values of $\alpha$ in Fig. \ref{fig:MixedModeExpt}a, for selected values of $\mathrm{CaCO}_{3}$ mass fraction $\bar{m}$. Unstable fast fracture accompanies the initiation of crack growth for all crack lengths and crack orientations. The values of ($K_I, K_{II}$) are normalised by the mode I fracture toughness for each cementation level (as reported in Section \ref{Sec:ModeIresults}). The failure locii in Fig. \ref{fig:MixedModeExpt}a are sensitive to the cementation level. This suggests that the damage zone is sufficiently large and diffuse at the point of instability that the initiation of crack growth cannot be reduced to the classical criterion of a critical value of $K_I^k$ and $K_{II}^k=0$ and $\Delta a=0$, as commonly assumed as for crack tip kinking in brittle solids \cite{Bazant1983,Bazant2002b}. However, in order to assess whether (\ref{eq:K_Ik}) and (\ref{eq:K_IIk}) can be used to reproduce the data of Fig. \ref{fig:MixedModeExpt}a, it is first necessary to determine an appropriate value of length $\Delta a$ for the putative kink at the crack tip. 

Our strategy is as follows. We determine a best-fitting value of $\Delta a$ such that the predicted value of crack kink direction  $\omega$  (such that $K_{II}^k=0$) is in agreement with the observed value of kink angle, over the full range of values of initial crack angle $\alpha$ and cementation level $\bar{m}$. Then, with this value of $\Delta a$ adopted as a material constant, the value of $K_{I}^k$ is determined as a function of $\alpha$ and $\bar{m}$. Classical kinking theory suggests that $K_{I}^k$ is invariant and equals the mode I fracture toughness of the solid, $K_{IC}$.

%For low $\bar{m}$ values, shear failure mechanisms operate. As shown in Fig. \ref{fig:ExptModeI}b, bio-cemented sands have a low $\sigma_f^c/\sigma_f^t$ ratio and this is accentuated at lower cementation levels. Accordingly, the resulting failure envelope would differ from conventional brittle rocks that fail under tension. Consistent with this, microscopy observations show that samples with low $\bar{m}$ exhibit shear failure modes with disaggregation at the grain scale. However, as the level of cementation increases, the fracture mechanisms change towards mode I cracking, with tensile cracks being observed along with cavities with dislocation of clusters of grains and %cementation.\\ 

A \textit{post-morten} examination of the samples was conducted to measure the sensitivity of observed kink angle $\omega$ to $\alpha$ and $\bar{m}$, see Fig. \ref{fig:MixedModeExpt}b. Predictions of $\omega$ are included by making use of Eq. (\ref{eq:K_IIk}) such that $K_{II}^k=0$ for $a/R=0.4$ and $S/R=0.8$. The theoretical estimates are obtained for selected values of $\Delta a /a$ in order to assess the role of the $T$-stress. The choice $\Delta a /a=0.01$ provides a good fit, and we note in passing that the choice $\Delta a /a=0.01$ corresponds to $\Delta a=D$, where $D$ is the particle diameter. This gives the putative crack length a clear physical interpretation and lies within the regime of $T$-stress dominance \cite{Liu2021b}. The need to include the role of $T$-stress in the prediction of kink angle is emphasised in Fig. \ref{fig:MixedModeExpt}d: the assumption of a negligible $T$-stress, or equivalently the assumption that ${\Delta a}=0$, underestimates the crack kinking angle as the degree of mode mix increases.  

%In addition, a theoretical estimate based on the $K_{II}=0$ crack growth criterion is included. Here, $T=0$ has been assumed: the role of the $T$-stress is quantified later on. The ($K_I, K_{II}$) trajectory labelled `$K_{II}^k=0$ criterion' is obtained by setting $K_{II}^k=0$ and $K_I^k=K_{Ic}$ in (\ref{eq:K_IIk}) and (\ref{eq:K_Ik}), respectively. The variation in stress intensity factor ratio ($K_I/K_{II}$) for the range of initial crack angles considered is shown in Fig. \ref{fig:MixedModeExpt}b, as per finite element analysis. 

We proceed to use Eq. (\ref{eq:K_Ik}) to predict the value of $K_I^k$ in each mixed-mode experiment, assuming that ($\Delta a/a=0.01$, see Fig. \ref{fig:MixedModeExpt}b). The results given in Fig. \ref{fig:MixedModeExpt}c show that the inferred value of $K_I^k$ is almost insensitive to the degree of mode mix. Further, the  $K_I^k$ values are close to the mode I fracture toughness at high $\bar{m}$ values. 
Some deviation is observed at low cementation levels, which is attributed to the diffuse nature of the cracking process \cite{Bazant1983,Bazant2002b} and the similar size of the initial flaw and the transition flaw size of the material. The low sensitivity of the inferred $K_I^k$ values to the degree of mode-mixity and their closeness to $K_{Ic}$ (for high values of $\bar{m}$) suggest that elastic-brittle fracture mechanics and a local mode I kinking criterion are appropriate for bio-cemented sands, particularly at high cementation levels. 

%Finally, Fig. \ref{fig:MixedModeTheory}c showcases the experimentally derived relationship between the phase angle $\psi=\tan^{-1} (K_{II}/K_I)$ and the initial crack angle $\alpha$. In agreement with the theory, smaller initial crack angles lead to an increase in the phase angle, as mode I conditions are approached. Also, no influence of the cementation is observed, with all $\bar{m}$ data points falling on top of each other for a given $\alpha$. This is also consistent with the theory as Eq. (\ref{eq:K_IIk}). 

\section{Discussion: A micromechanical model for particle fracture}
\label{Sec:Model}

\begin{figure}[H]
\begin{center}
\includegraphics[width=0.83\textwidth]{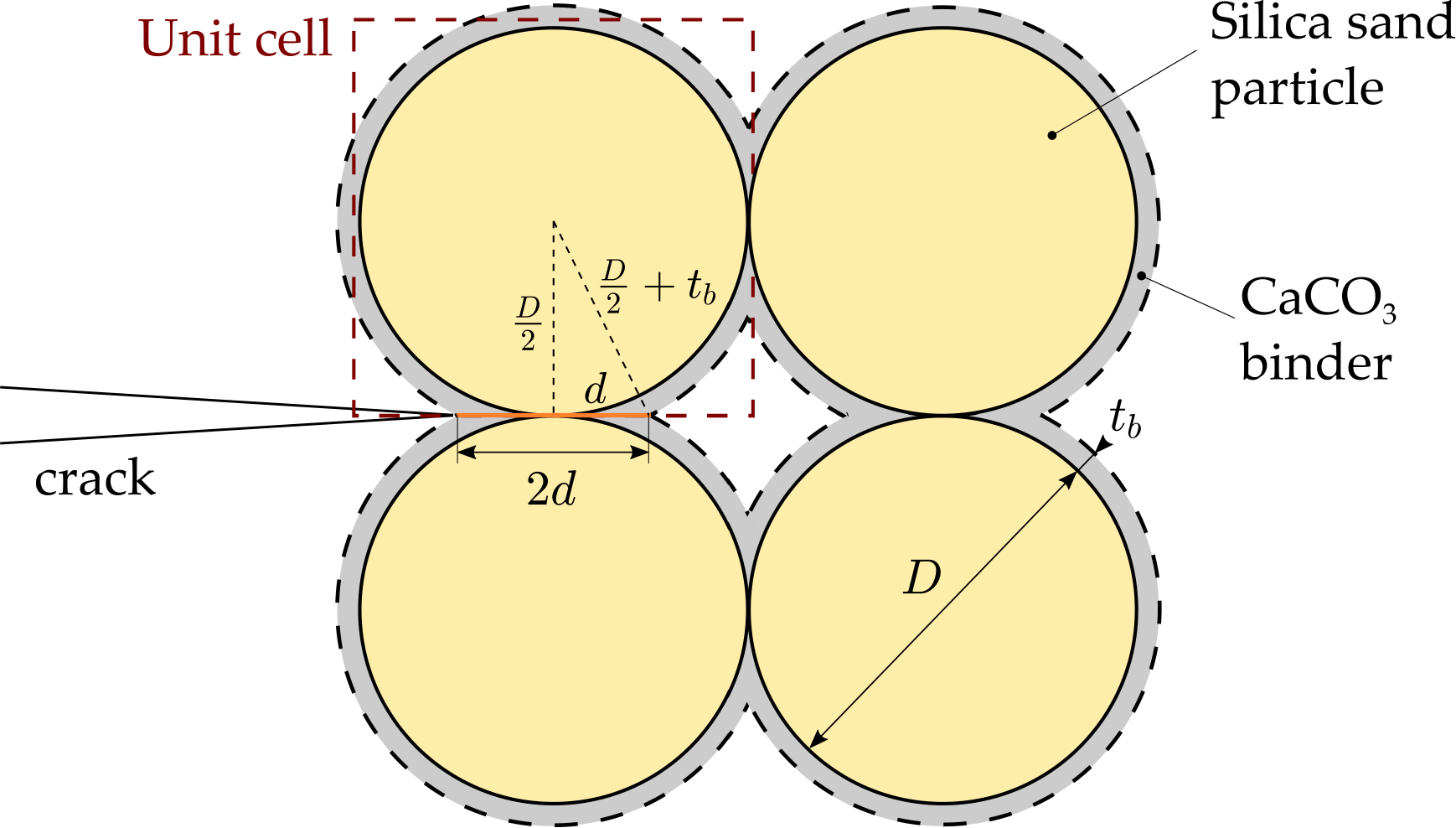}
\caption{A micromechanical description of fracture in bio-cemented sands. Silica sand particles of diameter $D$ are surrounded by a thin layer of binder $\mathrm{CaCO}_{3}$, of thickness $t_b$. The neck extends over a length $2d$.}
\label{fig:ModelUnitCell}
\end{center}
\end{figure}

A micromechanical model is now developed to predict the mode I fracture toughness of bio-mediated sandstones. Consider the idealised problem of a mode I $K$-field in the vicinity of the crack tip in a specimen that contains silica sand particles. The particles are treated as spheres of diameter $D$ and are coated by a layer of $\mathrm{CaCO}_{3}$ of thickness $t_b$, as shown in Fig. \ref{fig:ModelUnitCell}. The main purpose of the analysis is to develop scaling laws for fracture toughness and so the above idealisation of particle geometry is adequate for our purposes. Assume a simple cubic arrangement of sand particles, and restrict attention to a representative unit cell of side length $D$ at the tip of a long crack. The tensile load on a crack tip particle is related to the macroscopic stress at that location $\sigma^\infty$ according to 
\begin{equation} \label{Eq:PtoRemoteStress}
    P = \frac{\pi D^2 \sigma^\infty}{4 }
\end{equation}
Write the radius of the bonded contact between particles as $d$ and the average tensile stress acting over the bonded contact as $\sigma_L$, such that
\begin{equation} \label{Eq:LocalStress}
    P = \pi d^2 \sigma_L
\end{equation}
\noindent Consequently, the local contact stress $\sigma_L$ is related to the macroscopic stress $ \sigma^\infty$ by $\sigma_L = \left( D/2d \right)^2 \sigma^\infty$. 

Recall that the dominant term in the series expansion of mode I crack tip stress at a distance $r$ directly ahead of the crack tip is given by $\sigma = K_I /\sqrt{2\pi r} $. This stress field gives rise to the load $P$ over the unit cell directly ahead of the crack tip, where
\begin{equation} \label{Eq:Pintegral}
    P= \int_0^D \frac{ K_I}{\sqrt{2 \pi r}} D \, \text{d}r = \frac{2K_I D\sqrt{D}}{\sqrt{2 \pi}}
\end{equation}

Assume that crack advance occurs when the average contact stress $\sigma_L$ at the neck immediately ahead of the crack tip attains the critical value $\sigma_c$. Then, the plane strain fracture toughness $K_{Ic}$ follows from Eqs.  (\ref{Eq:LocalStress}) and (\ref{Eq:Pintegral}) as
\begin{equation} \label{Eq:Toughness1}
    K_{Ic} = \frac{\pi^{3/2}}{\sqrt{2}} \left( \frac{d}{D} \right)^2 D^{1/2} \sigma_c
\end{equation}

The mass fraction of binder is small and so the volume of binder can be approximated by $V_b \approx  \pi D^2 t_b$, along with $d^2 \approx D t_b$. Thus, the mass ratio $\bar{m}$ of binder to sand reads:
\begin{equation}
    \bar{m} = \frac{\textnormal{mass CaCO3}}{\textnormal{mass silica}} =  \frac{\rho_b}{\rho_s} \frac{6 d^2}{D^2} 
\end{equation}
where the subscripts s and b denote sand and binder, respectively, and ${\rho}$ denotes density.The above considerations suggest that the  mode I fracture toughness scales with the degree of cementation, and the tensile failure strength of the binder according  to:
\begin{equation}\label{Eq:Toughness2}
    K_{Ic}  =\frac{\pi^{3/2}}{\sqrt{2}}\frac{\bar{m} \rho_s}{6 \rho_b} D^{1/2} \sigma_c
\end{equation}

\begin{figure}[H]
  \makebox[\textwidth][c]{\includegraphics[width=0.8\textwidth]{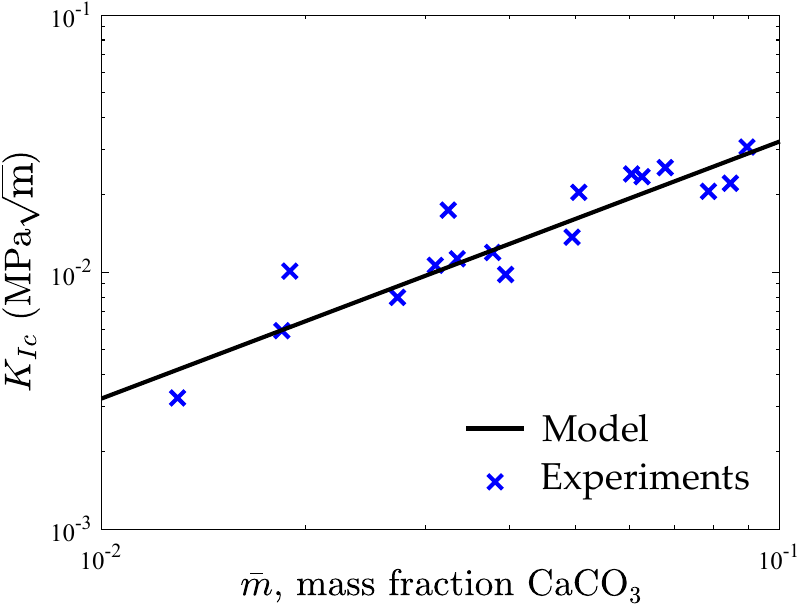}}%
  \caption{Micromechanical model predictions of fracture toughness $K_{Ic}$ versus degree of cementation, as given by the mass fraction of $\mathrm{CaCO}_{3}$.}
  \label{fig:ModelModeI}
\end{figure}

We proceed to investigate the ability of  Eq. (\ref{Eq:Toughness2}) to predict the sensitivity of fracture toughness to the degree of cementation. Assume that the density of silica sand and of $\mathrm{CaCO}_{3}$ binder equals $\rho_s=2.65$ Mg/m$^3$ and $\rho_b=2.71$ Mg/m$^3$, respectively. The typical particle has a diameter of $D\equiv D_{90}=0.3$ mm, and the local critical stress is taken to be $\sigma_c=29$ MPa. Close agreement between the predictions of the model and  experimental measurements is evident in Fig. \ref{fig:ModelModeI}, demonstrating the ability of the model to give the observed scaling of $K_{Ic}$ with $\mathrm{CaCO}_{3}$ mass fraction $\bar{m}$. 

%\noindent Now, for fitted value of $\sigma_c$ and an initial flaw size $a$, the strength reads,
%\begin{equation}
%    \sigma^\infty = \frac{K_{Ic} \left( f \right)}{\sqrt{\pi a}} = \frac{\frac{\pi^{3/2}}{\sqrt{2}}\frac{f \rho_b}{6 \rho_c} D^{1/2} \sigma_c}{\sqrt{\pi a}}
%\end{equation}

\section{Conclusions}
\label{Sec:Conclusions}

We have used microbiologically induced calcium carbonate precipitation (MICP) to generate bio-treated sandstone-like materials with a controlled degree of cementation. This enables us to shed light on the interplay between cementation and relevant material properties (porosity, permeability, tensile and compressive strengths). Moreover, the relationship between cementation and fracture behaviour is investigated. Mode I and mixed-mode fracture experiments are conducted and combined with theoretical analysis to gain mechanistic insight. A particle-level, micromechanical fracture model is developed to establish a correlation between the mode I fracture toughness $K_{Ic}$ and the mass fraction $\bar{m}$ of $\mathrm{CaCO}_{3}$. In addition, crack kinking theory is used to gain mechanistic insight into the role of $T$-stress upon the observed crack kinking angle. The main findings are:
\begin{itemize}
    \item Fracture toughness, tensile strength and compressive strength increase with cementation, while permeability and porosity decrease. Porosity is the main source of defects and consequently the ratio of compressive to tensile strength decreases with decreasing cementation as a result of the associated increase in porosity.
    \item For the range of $\bar{m}$ values considered, bio-cemented sands exhibit low values of fracture toughness and strength relative to other rock-like materials, with only mudstones providing similar strengths. The critical defect size of the bio-cemented sands is $a_0=1$ mm, which is comparable to that of shale rock. 
    \item The measured crack kinking angles increase with the degree of mode-mixity and can be adequately captured by including the role of the $T$-stress.
    \item The mode I kink tip fracture toughness $K_I^k$ can be inferred from the mixed-mode experiments for all initial crack angles. At low cementation values, microcracking is more diffuse and linear elastic fracture mechanics does not apply for the specimen size adopted in the fracture tests. 
    \item A micromechanical model for fracture toughness is supported by the experimental data, demonstrating its ability to capture the sensitivity of fracture toughness to the degree of cementation.    
\end{itemize}

\section{Acknowledgments}
\label{Sec:Acknowledge of funding}

The authors acknowledge funding from the International Centre for Advanced Materials (ICAM) (ICAM39: 2016-2020). E. Mart\'{\i}nez-Pa\~neda was supported by an UKRI Future Leaders Fellowship [grant MR/V024124/1] and additionally acknowledges financial support from the Royal Commission for the 1851 Exhibition through their Research Fellowship programme (RF496/2018). N. A. Fleck is grateful for financial support from the ERC Advanced Grant MULTILAT, grant no. 669764.


%\newpage
%\appendix
%

%\processdelayedfloats % This is basically to include the figures here, before the appendix



\end{document}